\documentclass[twocolumn,showpacs,preprintnumbers,amsmath,amssymb]{revtex4}

\usepackage{graphicx}
\usepackage{dcolumn}
\usepackage{bm}
\usepackage{amsmath}

\begin{document}


\title{Identified charged antiparticle 
to particle ratios near midrapidity in Cu+Cu collisions 
at $\sqrt{s_{_{\it NN}}}$ = 62.4 and 200 GeV}

\author{
%
%
B.Alver$^4$,
B.B.Back$^1$,
M.D.Baker$^2$,
M.Ballintijn$^4$,
D.S.Barton$^2$,
R.R.Betts$^6$,
R.Bindel$^7$,
W.Busza$^4$,
Z.Chai$^2$,
V.Chetluru$^6$,
E.Garc\'{\i}a$^6$,
T.Gburek$^3$,
K.Gulbrandsen$^4$,
J.Hamblen$^8$,
I.Harnarine$^6$,
C.Henderson$^4$,
D.J.Hofman$^6$,
R.S.Hollis$^6$,
R.Ho\l y\'{n}ski$^3$,
B.Holzman$^2$,
A.Iordanova$^6$,
J.L.Kane$^4$,
P.Kulinich$^4$,
C.M.Kuo$^5$,
W.Li$^4$,
W.T.Lin$^5$,
C.Loizides$^4$,
S.Manly$^8$,
A.C.Mignerey$^7$,
R.Nouicer$^2$,
A.Olszewski$^3$,
R.Pak$^2$,
C.Reed$^4$,
E.Richardson$^7$,
C.Roland$^4$,
G.Roland$^4$,
J.Sagerer$^6$,
I.Sedykh$^2$,
C.E.Smith$^6$,
M.A.Stankiewicz$^2$,
P.Steinberg$^2$,
G.S.F.Stephans$^4$,
A.Sukhanov$^2$,
A.Szostak$^2$,
M.B.Tonjes$^7$,
A.Trzupek$^3$,
G.J.van~Nieuwenhuizen$^4$,
S.S.Vaurynovich$^4$,
R.Verdier$^4$,
G.I.Veres$^4$,
P.Walters$^8$,
E.Wenger$^4$,
D.Willhelm$^7$,
F.L.H.Wolfs$^8$,
B.Wosiek$^3$,
K.Wo\'{z}niak$^3$,
S.Wyngaardt$^2$,
B.Wys\l ouch$^4$\\
\vspace{3mm}
\small
%
%
%
%
$^1$~Argonne National Laboratory, Argonne, IL 60439-4843, USA\\
$^2$~Brookhaven National Laboratory, Upton, NY 11973-5000, USA\\
$^3$~Institute of Nuclear Physics PAN, Krak\'{o}w, Poland\\
$^4$~Massachusetts Institute of Technology, Cambridge, MA 02139-4307, USA\\
$^5$~National Central University, Chung-Li, Taiwan\\
$^6$~University of Illinois at Chicago, Chicago, IL 60607-7059, USA\\
$^7$~University of Maryland, College Park, MD 20742, USA\\
$^8$~University of Rochester, Rochester, NY 14627, USA\\
}

\begin{abstract} 
\noindent
Antiparticle to particle ratios for identified protons, kaons and 
pions at $\sqrt{s_{_{NN}}}$ = 62.4 and 200 GeV in Cu+Cu collisions 
are presented as a function of centrality for the midrapidity region 
of  $0.2 < \eta < 1.4$.  No strong dependence on centrality is observed. 
For  the $\langle \overline{p} \rangle / \langle p \rangle$ ratio at 
$\langle p_T \rangle \approx 0.51$ GeV/c,  we observe an average value 
of  $0.50 \pm 0.003_\textrm{(stat)} \pm 0.04_\textrm{(syst)}$ and 
$0.77 \pm 0.008_\textrm{(stat)} \pm 0.05_\textrm{(syst)}$  for the 
10\% most central collisions of 62.4 and 200 GeV Cu+Cu, respectively.  
The values for all three particle species measured at 
$\sqrt{s_{_{NN}}}$ = 200 GeV are in agreement within systematic 
uncertainties with that seen in both heavier and lighter systems 
measured at the same RHIC energy. This indicates that system size does 
not appear to play a strong role in determining the midrapidity 
chemical freeze-out properties affecting the antiparticle to particle ratios
of the three most abundant particle 
species produced in these collisions.
\end{abstract}


\maketitle

This paper reports the first measurement of antiparticle to particle 
ratios of pions, kaons and protons in Cu+Cu collisions at  
$\sqrt{s_{_{NN}}}$ = 62.4 and 200 GeV.  The data were taken during the
2005 run using the PHOBOS detector at the Relativistic Heavy Ion Collider 
(RHIC) at Brookhaven National Laboratory. Antiparticle to particle 
ratio measurements are a useful probe in the context of understanding 
the chemical freeze-out properties of the created state of matter. 
The antiproton to proton ratios, in particular, represent a direct 
measure of the extent to which the central collision zone is baryon 
free, and the data also provide additional insight into baryon transport
in these collisions. A primary focus of this article is a comparison of results
obtained in the  
Cu+Cu collision system to both smaller and larger systems in order to 
determine the effect of system size on the measurement. 

The strength of this experimental 
result stems primarily from the 
fact that all effects of acceptance and efficiency cancel
in the PHOBOS  measurement of identified particle ratios.
In addition, the excellent collision vertex resolution 
allows for a tight distance-of-closest-approach selection of identified 
particle tracks, which reduces contributions from 
weak decays and other sources of secondary particles.

Results presented here are obtained using the PHOBOS two-arm 
spectrometer \cite{phobos2}.   The active elements of the tracking 
detectors in the spectrometer are constructed of highly segmented 
silicon pad sensors, with the energy deposited in each pad recorded. 
Each spectrometer arm has a geometrical acceptance of $\pm 0.1$ radians 
in the azimuthal angle and 0.48 to 1.37 radians in polar angle  
($0.2 < \eta < 1.4$).  The outer 9 layers of each 15 layer 
spectrometer arm are located in a 2~T magnetic field provided by 
the dipole magnet. The magnetic field is  perpendicular to the  
plane of the spectrometer. 
For a given field setting, oppositely charged particles 
bend in opposite directions in the 
plane of the spectrometer.
The direction of the magnetic 
field is reversed frequently during the course of data taking, 
recording  approximately the same statistics for each field setting 
and allowing for a dataset that further reduces systematic uncertainties arising 
from non-uniform beam conditions that occur in the long RHIC runs. 
Further details of the PHOBOS detector setup can be found in 
Refs.~\cite{phobos1, phobos2, whitepaper}. 

Due to the geometrical asymmetry of a given spectrometer arm in 
$\eta$-coverage, charged particles bending in opposite directions 
have different acceptances.  Hence, to obtain the final ratios, 
opposite magnetic field settings for particles bending in the same 
direction in the same spectrometer arm are used so that they have 
the same acceptance. 
Denoting positively and negatively charged particles
$\mathrm{h^+}$ and $\mathrm{h^-}$, at a given magnet polarity
setting of $\mathrm{B^+}$ or $\mathrm{B^-}$, the formula for 
calculating the particle ratios for forward bending tracks in a 
given spectrometer arm for $N_\mathrm{h}$ particles and 
$N_\mathrm{events}$ events is
\[
\frac{\langle \mathrm{h^-} \rangle}{\langle \mathrm{h^+} \rangle} =
\frac{N^\mathrm{B^+}_\mathrm{h^{-}}(p_{T},\mathrm{Centrality}) 
\times N^\mathrm{B^-}_\mathrm{events}(\mathrm{Centrality}) }
{N^\mathrm{B^-}_\mathrm{h^{+}}(p_{T},\mathrm{Centrality}) 
\times N^\mathrm{B^+}_\mathrm{events}(\mathrm{Centrality})}.
\]
\noindent
In a similar manner, the formula for backward bending tracks in 
a given spectrometer arm is given by switching the magnet polarity setting.  
Four independent like-particle ratios are obtained, and the data presented are 
the statistically averaged results of these four measurements. 

\begin{figure}
\centerline{
\includegraphics[width=0.99\columnwidth]{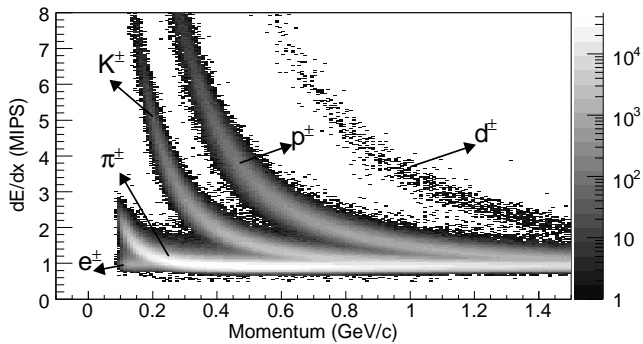}}
\caption{Energy deposited by the reconstructed tracks in both 
PHOBOS spectrometer arms as a function of momentum 
for  $\sqrt{s_{_{NN}}}$ = 200 GeV Cu+Cu collisions.  The bands are 
labeled for different particle species, and the current analysis 
is for pions, kaons and protons. \label{Fig:PID}}
\end{figure}

\begin{figure}
\centering
\includegraphics[width=0.99\columnwidth]{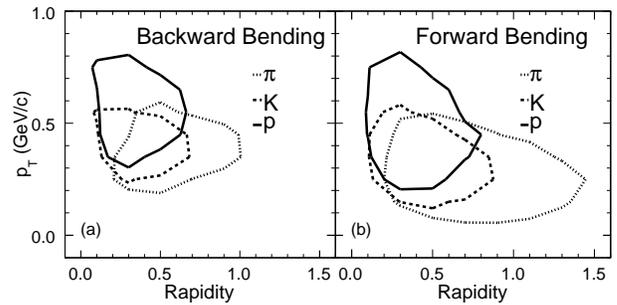}
\caption{Acceptance of identified particles in transverse momentum 
and rapidity space for $\sqrt{s_{_{NN}}}$ = 200 GeV Cu+Cu collisions 
using the PHOBOS  spectrometer. Backward Bending (a): Acceptance
of particle bending away from the beam. Forward Bending (b): The acceptance 
of the particles bending towards the beam. Lines represent the 
20\% acceptance contour level.
\label{Fig:Acceptance}}
\end{figure}

There are two steps involved in the extraction of identified particles 
in the spectrometer: tracking of the particles to  obtain their 
momentum, and measurement of the energy deposited  in the silicon 
to enable particle identification at a given momentum value.  
Details of the track reconstruction algorithm are described 
in~\cite{phobos130}.  Only particles that traversed the full 
spectrometer arm are included in this analysis.  The  
track selection is based on an upper limit of 0.35~cm 
distance-of-closest-approach (DCA) and a $\chi^2$ probability 
requirement on the tracking fit, which allows for rejection of 
tracks with incorrectly assigned hits and thereby improves momentum 
and particle identification resolution.  Particle identification 
(PID) is based on measuring the truncated  mean of the specific 
ionization, $dE/dx$, observed in the silicon  detectors, as a 
function of momentum.  The results of this are illustrated in  
Fig.~\ref{Fig:PID} for Cu+Cu collisions at $\sqrt{s_{_{NN}}}$ 
= 200 GeV. A projection of the $dE/dx$ distribution for a given  
momentum bin is analyzed to produce  PID bands. The choice of 
momentum bin size is driven by the available  statistics.  
Local maxima are fitted with Gaussians, from which the mean  
and sigma are obtained as a function of momentum for each species band.  
Valid PID bands, versus momentum, are defined as $\pm2\sigma$ from 
the mean.  The upper limit of particle separation
in momentum is set at the intersection of the 3$\sigma$ bands, 
thus ensuring a negligible contamination ($< 0.1$\%). 
The corresponding acceptance regions for identified particles in 
transverse momentum, $p_T$, and rapidity, $y$, are shown in  
Fig.~\ref{Fig:Acceptance}.   As a cross-check, the PID bands 
were also determined using a modified Bethe-Bloch function 
technique, as performed in prior PHOBOS measurements
\cite{phobos130,PHOBOSAuAu200Ratios,phobosdAratios,PHOBOSPPRatios}. 
Both approaches yield results consistent to better than 1\% .

The primary  hardware trigger for the Cu+Cu data  was 
similar to that used for much of the earlier PHOBOS Au+Au data, 
and required at least two charged particle hits in each of the 
symmetrically located   Paddle scintillator counters 
(3.2$<$$|\eta|$$<$4.5) within a time-difference of $\Delta t$$<$10 ns.  
This trigger ensures no loss of central collision events that can 
occur if the primary hardware trigger is based solely on the 
zero-degree calorimeters (ZDCs).  
To enhance the statistics of data near the 
nominal vertex position ($z_{\rm vtx}$ = 0) and 
maximize the acceptance in the spectrometer, PHOBOS also employed two 
symmetrically located rings of 10 \v{C}erenkov counters 
(4.4$<$$|\eta|$$<$4.9) that enabled a fast $z$-vertex 
determination.

The offline event selection utilized calibrated Paddle 
signals with time differences of  $\Delta t \leq 5$ ns and 
only events with  a reconstructed collision vertex of 
$|z_{\rm vtx} |\leq 9$~cm. The additional off-line 
$z$-vertex selection is applied to ensure good tracking efficiency 
and a consistent acceptance. The vertex reconstruction was 
optimized for the lower  multiplicity Cu+Cu data through 
the development of two new  vertex reconstruction techniques.  
The first technique reconstructed a high-efficiency 
low-resolution vertex position from the single-layered
Octagon detector \cite{Garcia:OctVertexPaper} and the second 
reconstructed a lower-efficiency but higher-resolution vertex 
position by  utilizing two-hit tracklets in the Vertex detector 
and the first two layers of the spectrometer.  
A simultaneous valid vertex reconstruction in both methods was 
required in order to maximize data purity. 
This event selection is similar to that used in Ref.~\cite{phobosCuCu_pT}.  
In addition, the data is classified into different subsets 
based on the  run-wise $\langle x \rangle$ and $\langle y \rangle$ values of 
the reconstructed collision point ($z$-axis is parallel to the beam).  As the 
spectrometer has a very small azimuthal acceptance, the tracking 
has some sensitivity to the $x$ and $y$ positions of the collision. 
Data subsets for different polarities were matched based on 
this ``beam orbit'' classification and  subsequently used for
determination of the particle ratios.  

The collision centrality is defined through bins of fractional 
total inelastic cross section, where the most central bin covering 
the 3\% most central events is defined as 0-3\%.  
Current results are reported down to the mid-central bin of 
40-45\%, with the limit set primarily by the desire for sufficient 
kaon statistics to enable a robust measurement.  
In order to divide the data into bins of fractional cross section, 
it is necessary to  understand the trigger and vertex 
reconstruction efficiency of the PHOBOS detector for Cu+Cu collisions.  
Two methods to  extract this efficiency were used to determine a total 
efficiency (including vertexing) of $75\pm5$\% and $84\pm5$\% for the 62.4 
and 200 GeV Cu+Cu data, respectively.  
Both methods have been successfully employed for a wide range of 
past data covering Au+Au collisions from 19.6 to 200 GeV 
\cite{phobos:tracklet200_20} as well as the much lower multiplicity 
d+Au collision data at 200 GeV \cite{phobos:dAu_dNdEta}.  
It is important to note that the inefficiency is located in 
peripheral data and there is no inefficiency  for the 
reported centrality bins.   
The data utilize the centrality classes as determined  from the 
total energy deposited in the silicon Octagon detector \cite{phobosCuCu_pT}.  
The corresponding average number of participating nucleons, 
$\langle N_{\rm part} \rangle$, for each centrality class
are calculated as detailed in Refs.~\cite{phobosCuCu_pT,whitepaper}.

In order to obtain ratios of the primary yields, the inclusive 
(measured) particle ratios are corrected for the asymmetric 
absorption of antiparticles versus particles in the detector 
materials, contamination by secondary particles, and feed-down 
from hyperon decay.  The methods of obtaining the correction 
factors, which are applied directly to the measured ratios, 
are similar to those used in prior analysis \cite{phobosdAratios}, now 
calculated for Cu+Cu data. The next three paragraphs 
summarize the techniques and magnitude of the corrections 
for the $\sqrt{s_{_{NN}}} = 200$ GeV data in Cu+Cu collisions.  
Average values for the corrections to the 62.4 GeV data are similar or smaller.

The absorption correction is obtained using a GEANT simulation of 
the PHOBOS beam-pipe coupled to the spectrometer acceptance. 
The percentage of absorbed yield is obtained for each species 
as a function of transverse momentum, and the final correction 
is obtained by taking the average of two different hadronic 
interaction packages, Gheisha and Fluka~\cite{GEANT}. 
Proton ratios have the largest absorption correction, 
mainly due to increased annihilation of antiprotons as 
compared to protons. The absorption correction to the ratios, 
averaged over the accepted transverse momentum, are 
$\sim 1$\%, $<0.05$\% and $\sim 5$\% for pions, kaons 
and protons, respectively. 

We define secondaries as the yield of particles produced from both 
the beam-pipe and the PHOBOS detector material. 
Events from HIJING~\cite{HIJING} are used to simulate all particles 
produced in the collision, which are transported using 
a full GEANT simulation of the detector to obtain the correction. 
The secondary correction to the ratios, averaged over the 
acceptance, is found to be $\sim 2$\%, $\sim 0$\% and 
$\sim 1$\% for pions, kaons and protons, respectively.

Feed-down particles produced from the decay of hyperons 
contribute to the non-primary yield, and thus must be 
corrected for in order to obtain the primary particle ratios.  
The proton ratios are more sensitive to weak 
decays as compared to pions and kaons \cite{phobos130}.  
The pion and kaon corrections to the individual yields 
essentially cancel in the final ratios, thus no correction is applied.  
However, the protons are more complicated. This is illustrated by the 
fact, known from early RHIC results, that 
the $\overline{\Lambda}/\Lambda$ and $\overline{p}/p$ ratios 
are not unity in Au+Au collisions at 
RHIC energies \cite{PHENIXAuAu130lambda,STARAuAu130pbarp}.  
Comparisons between HIJING calculations and data for protons 
and lambdas in Au+Au collisions have indicated differences 
in $\overline{\Lambda}/\overline{p}$ and $\Lambda/p$ ratios 
of up to factors of three, with HIJING being lower than 
data \cite{PHOBOSAuAu200Ratios}. 
Within PHOBOS, due to the excellent resolution of the $z$-vertex 
reconstruction ($\sigma_z \le 0.25$ cm), a requirement of DCA 
less than $0.35$ cm between the reconstructed tracks and event 
vertex is used, which removes a significant fraction of weak 
decays \cite{PHOBOS62AuAuPID}.  
For the new Cu+Cu data, HIJING is used to obtain a baseline 
expectation for $\overline{\Lambda}/\overline{p}$ and 
$\Lambda/p$ and then these ratios are varied by up to a factor of three.  
In addition, the calculations for Cu+Cu were manually 
adjusted with factors that reproduced measured values 
for protons and lambdas in Au+Au collisions.  
This work resulted in an average value used to obtain a 
final $\Lambda$ and $\overline{\Lambda}$ feed-down 
correction for the proton ratios of $\sim 2.2$\%.  
The known variations between data and simulation for 
Au+Au is included in the systematic uncertainties on the proton 
ratio feed-down correction for Cu+Cu. 

\begin{figure}[!th]
\centerline{\includegraphics[width=0.99\columnwidth]{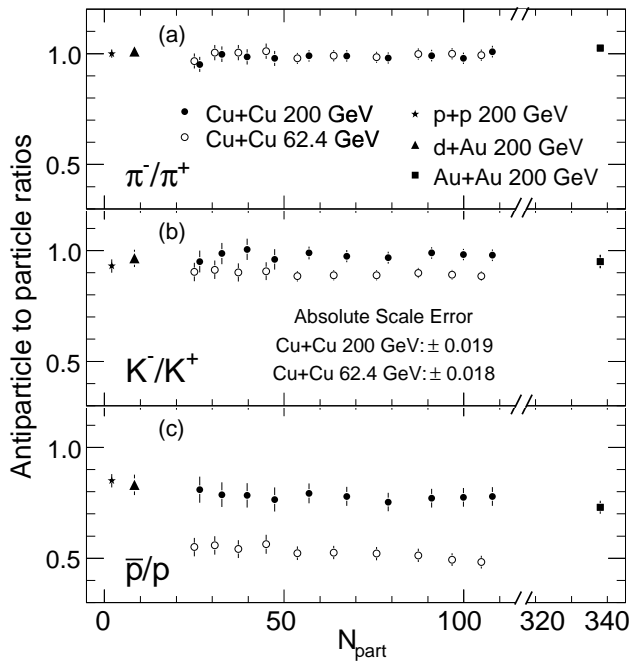}}
\caption{Antiparticle to particle ratios in Cu+Cu collisions, as a 
function of the number of participants, for pions (a), kaons (b) and protons (c). 
Open (closed) circles represent $\sqrt{s_{_{\it NN}}} = 62.4$ GeV
(200 GeV) data. 
The error bars represent the combined (1$\sigma$) statistical 
and systematic uncertainties, and an additional scale error is 
indicated on the figure. PHOBOS data for p+p 
{\protect\cite{PHOBOSPPRatios}}, minimum-bias d+Au  
{\protect\cite{phobosdAratios}}, and central Au+Au 
{\protect\cite{PHOBOSAuAu200Ratios}} collisions at 
$\sqrt{s_{_{\it NN}}} = 200$ GeV are also shown.  
 \label{Fig:ratio_centrality}
}
\end{figure}

The final step in data analysis involves determination of systematic 
uncertainties, which arise from event selection, PID cuts, and the 
three correction factors.  
An additional contribution comes from repeating the analysis for 
different subsets of data based on the aforementioned  
$\langle x \rangle$ and $\langle y \rangle$ collision vertex 
position classifications. No single uncertainty (parameter) 
dominates the final systematic error, typically the smallest 
contribution comes from the PID cuts and the largest from either 
the event selection or, in the case of the proton ratios, the 
feed-down correction. The final systematic uncertainty for a 
given centrality is determined from the statistically weighted 
average of the uncertainty determined for each parameter for 
different arms and bending directions. 
A thorough investigation of the track selection $\chi^2$ probability 
cut has shown a variation independent of the species and arm, 
but dependent on the bending direction. 
Hence, this effect yields a scale systematic uncertainty that, 
for each collision energy, is independent of both centrality 
and particle species.

The measured primary antiparticle to particle ratios for Cu+Cu 
collisions at $\sqrt{s_{_{NN}}}$ = 62.4 and 200 GeV as a function 
of the number of participants, $N_{\rm part}$, are shown in 
Fig.~\ref{Fig:ratio_centrality}. The data, also given in 
Tables \ref{Table:CuCu63} and \ref{Table:CuCu200},
are averaged over the acceptance as illustrated in Fig.~\ref{Fig:Acceptance}. 
No strong centrality dependence of the proton, kaon or pion 
antiparticle to particle ratios is observed. 
The particle ratios measured in Cu+Cu follow similar trends 
with collision energy as observed in Au+Au.  
In Cu+Cu, the pion particle ratios are consistent with unity 
at both 62.4 and 200 GeV, the kaon ratios have a weak collision 
energy dependence but reach unity at 200 GeV, and the proton 
ratios show the strongest variation, rising from an average value of 
$\langle \overline{p} \rangle/\langle p \rangle = 0.50\pm0.003\pm0.04$ 
at 62.4 GeV to $0.77\pm0.008\pm0.05$ at 200 GeV, 
for the 10\% most central collisions.  
At the same energy of $\sqrt{s_{_{NN}}} = 200$ GeV, the results 
for particle ratios of pions, kaons and protons in Cu+Cu collisions 
are consistent, within uncertainties, to those found by PHOBOS for 
p+p \cite{PHOBOSPPRatios}, d+Au \cite{phobosdAratios} 
and  central Au+Au \cite{PHOBOSAuAu200Ratios} collisions,
as shown in Fig.~\ref{Fig:ratio_centrality}.  
However, the current PHOBOS data does allow
for a slight decrease in the 
$\langle \overline{p} \rangle/\langle p \rangle$ ratio as a 
function of increasing $\langle N_{\rm part} \rangle$ across 
the different systems.  
Only a single minimum-bias point is shown for d+Au
since, as discussed in Ref.~\cite{phobosdAratios},
the more appropriate variable to study the centrality 
dependence of particle ratios in this very asymmetric 
system is the number of collisions per deuteron participant ($\nu$).
Similar to the present data, no centrality dependence was found
in d+Au, even for the $\langle \overline{p} \rangle/\langle p \rangle$
ratio over a range of a factor of four in $\nu$ and up to 
a total number of participants larger than 
fifteen \cite{phobosdAratios}. 
The minimum-bias point in d+Au is formed by a weighted average using
combined statistical and systematic uncertainities
and the associated value of $\langle N_{\rm part} \rangle$
= 8.3 is given in 
Ref.~\cite{PHOBOSdAuNPart}. 

\begin{figure}[!t]
\centering
\includegraphics[width=0.99\columnwidth]{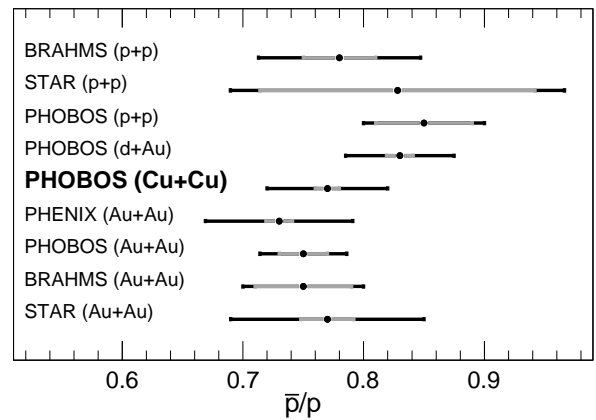}
\caption{Antiproton to proton ratios for 
$\sqrt{s_{_{NN}}} = 200$ GeV collisions 
from RHIC, compared to the new result for central Cu+Cu collisions (bold).  
The gray bar represents the statistical error, and  the black bar 
represents the combined statistical and systematic uncertainty. 
Data and references are listed in Table {\protect \ref{Table:CuCu200}}.  
\label{Fig:CompareRatios_graphic}}
\end{figure}

\begin{table*}[tb]
\caption{
\label{Table:CuCu63}
Cu+Cu collision results for antiparticle to particle ratios at 
$\sqrt{s_{_{NN}}}$ = 62.4 GeV.  
The average transverse momenta of each result is 
$\langle p_T \rangle \approx$ 0.31, 0.36 and 0.50 GeV/c for pions, 
kaons and protons, respectively.  
The uncertainties on $\langle N_{\rm part} \rangle$ are 90\% C.L.\ systematic 
and on the ratios are standard (1$\sigma$) statistical and systematic, respectively.  
There is an additional absolute scale systematic uncertainty of $\pm$0.018 on 
all particle ratio values.  
Average particle ratios, including the scale systematic uncertainty,  
for the 10\% most central collisions are 
$1.00\pm0.001\pm0.03$, $0.89\pm0.004\pm0.03$ 
and $0.50\pm0.003\pm0.04$ for pions, kaons and protons, respectively.}
\begin{ruledtabular}
\begin{tabular}{ccccc}
Centrality & $\langle N_{\rm part} \rangle$ & 
$\langle \pi^- \rangle / \langle \pi^+ \rangle$ & 
$\langle K^- \rangle/ \langle K^+ \rangle$ &
$\langle \bar{p} \rangle/ \langle p \rangle$\\ \hline 
$0 - 3$ \% & $106\pm3$  &  $0.99\pm0.002\pm0.03$ &  $0.89\pm0.006\pm0.02$ &  $0.48\pm0.005\pm0.03$\\
$3 - 6$ \% & $ 97\pm3$  &  $1.00\pm0.002\pm0.03$ &  $0.89\pm0.006\pm0.02$ &  $0.49\pm0.005\pm0.03$\\
$6 - 10$ \% & $ 88\pm3$  &  $1.00\pm0.002\pm0.03$ &  $0.90\pm0.007\pm0.02$ &  $0.51\pm0.005\pm0.03$\\
$10 - 15$ \% & $ 76\pm3$  &  $0.98\pm0.002\pm0.03$ &  $0.89\pm0.007\pm0.02$ &  $0.52\pm0.006\pm0.03$\\
$15 - 20$ \% & $ 65\pm3$  &  $0.99\pm0.002\pm0.03$ &  $0.89\pm0.008\pm0.02$ &  $0.53\pm0.006\pm0.03$\\
$20 - 25$ \% & $ 55\pm3$  &  $0.98\pm0.002\pm0.03$ &  $0.88\pm0.009\pm0.02$ &  $0.52\pm0.007\pm0.03$\\
$25 - 30$ \% & $ 47\pm3$  &  $1.01\pm0.003\pm0.04$ &  $0.91\pm0.011\pm0.04$ &  $0.56\pm0.008\pm0.04$\\
$30 - 35$ \% & $ 38\pm3$  &  $1.00\pm0.003\pm0.04$ &  $0.90\pm0.012\pm0.04$ &  $0.54\pm0.009\pm0.04$\\
$35 - 40$ \% & $ 32\pm3$  &  $1.00\pm0.003\pm0.04$ &  $0.91\pm0.014\pm0.04$ &  $0.56\pm0.010\pm0.04$\\
$40 - 45$ \% & $ 26\pm3$  &  $0.97\pm0.004\pm0.04$ &  $0.90\pm0.017\pm0.04$ &  $0.55\pm0.012\pm0.04$\\
\end{tabular}
\end{ruledtabular}
\end{table*}

\begin{table*}[htb]
\caption{
\label{Table:CuCu200}
Top: New Cu+Cu collision results for 
antiparticle to particle ratios at $\sqrt{s_{_{NN}}}$ = 200 GeV.  
The average transverse momenta of each result is 
$\langle p_T \rangle \approx$ 0.31, 0.37 and 0.51 GeV/c for pions, 
kaons and protons, respectively.  
The given uncertainties, as well as an additional scale 
systematic of $\pm$0.019 on all Cu+Cu ratios, are as described 
in Table {\protect\ref{Table:CuCu63}}. 
Bottom: Published midrapidity results for p+p, minimum-bias (minbias) d+Au 
and central Au+Au collisions at 200 GeV, as well as the 
result (including the scale systematic) for the 10\% most 
central collisions in Cu+Cu at 200 GeV. 
Values are rounded to the shown precision. 
The STAR (p+p) values for pions and kaons have only the total error.}
\begin{ruledtabular}
\begin{tabular}{ccccc}
Centrality & $\langle N_{\rm part} \rangle$ & 
$\langle \pi^- \rangle / \langle \pi^+ \rangle$ & 
$\langle K^- \rangle/ \langle K^+ \rangle$ &
$\langle \bar{p} \rangle/ \langle p \rangle$\\ \hline 
 $0 - 3$ \%  &  $108\pm3$  &   $1.01\pm0.003\pm0.03$& $0.98\pm0.012\pm0.02$& $0.78\pm0.013\pm0.04$\\
 $3 - 6$ \% &  $ 101\pm3$  &   $0.98\pm0.003\pm0.03$& $0.98\pm0.012\pm0.02$& $0.77\pm0.013\pm0.04$\\
 $6 - 10$ \%  &  $ 91\pm3$  &   $0.99\pm0.003\pm0.03$& $0.99\pm0.013\pm0.02$& $0.77\pm0.014\pm0.04$\\
 $10 - 15$ \% &  $ 79\pm3$  &   $0.98\pm0.003\pm0.03$& $0.97\pm0.014\pm0.02$& $0.75\pm0.015\pm0.04$\\
 $15 - 20$ \% &  $ 67\pm3$  &   $0.99\pm0.003\pm0.03$& $0.97\pm0.016\pm0.02$& $0.78\pm0.016\pm0.04$\\
 $20 - 25$ \% &  $ 57\pm3$  &   $0.99\pm0.004\pm0.03$& $0.99\pm0.018\pm0.02$& $0.79\pm0.018\pm0.04$\\
 $25 - 30$ \% &  $ 48\pm3$  &   $0.98\pm0.004\pm0.04$& $0.96\pm0.019\pm0.04$& $0.77\pm0.019\pm0.05$\\
 $30 - 35$ \% &  $ 40\pm3$  &   $0.99\pm0.005\pm0.04$& $1.01\pm0.023\pm0.04$& $0.78\pm0.021\pm0.05$\\
 $35 - 40$ \% &  $ 33\pm3$  &   $1.00\pm0.005\pm0.04$& $0.99\pm0.025\pm0.04$& $0.79\pm0.024\pm0.05$\\
 $40 - 45$ \% &  $ 27\pm3$  &   $0.95\pm0.006\pm0.04$& $0.95\pm0.029\pm0.04$& $0.81\pm0.028\pm0.05$\\ \hline
\multicolumn{1}{l}{Experiment (system)~[Ref.]} & Centrality &
$\langle \pi^- \rangle / \langle \pi^+ \rangle$ & 
$\langle K^- \rangle/ \langle K^+ \rangle$ &
$\langle \bar{p} \rangle/ \langle p \rangle$\\ \hline 
\multicolumn{1}{l}{BRAHMS (p+p)~{\protect\cite{BRAHMSPPRatios}}} & -- & $1.02\pm0.01~\pm0.07$ & $0.97\pm0.05~\pm0.07$ & $0.78\pm0.03~\pm0.06$\\
\multicolumn{1}{l}{STAR (p+p)~{\protect\cite{STARPPRatios}}} & -- & $0.99\pm0.14$ & $0.94\pm0.08$ & $0.83\pm0.114\pm0.08$\\
\multicolumn{1}{l}{PHOBOS (p+p)~{\protect\cite{PHOBOSPPRatios}}} & -- & $1.00\pm0.012\pm0.02$ & $0.93\pm0.05~\pm0.03$ & $0.85\pm0.04~\pm0.03$\\
\multicolumn{1}{l}{PHOBOS (d+Au)~{\protect\cite{phobosdAratios}}} & minbias & $1.01\pm0.004\pm0.02$ & $0.96\pm0.014\pm0.04$ & $0.83\pm0.011\pm0.05$\\
\multicolumn{1}{l}{PHOBOS (Cu+Cu)} & $0 - 10$ \% & $0.99\pm0.002\pm0.03$ & $0.98\pm0.007\pm0.03$ & $0.77\pm0.008\pm0.05$ \\
\multicolumn{1}{l}{PHENIX (Au+Au)~{\protect\cite{PHENIXAuAu200Ratios}}} & $0 - 5$ \% & $0.98\pm0.004\pm0.06$ & $0.93\pm0.007\pm0.05$ &  $0.73\pm0.011\pm0.06$\\
\multicolumn{1}{l}{PHOBOS (Au+Au)~{\protect\cite{PHOBOSAuAu200Ratios}}} & $0 - 12$ \% & $1.03\pm0.006\pm0.02$ & $0.95\pm0.03~\pm0.03$ & $0.73\pm0.02~\pm0.03$\\
\multicolumn{1}{l}{BRAHMS (Au+Au)~{\protect\cite{BRAHMSAuAu200Ratios}}} & $0 - 20$ \% & $1.01\pm0.010\pm0.04$ & $0.95\pm0.05~\pm0.04$ & $0.75\pm0.04~\pm0.03$\\
\multicolumn{1}{l}{STAR (Au+Au)~{\protect\cite{STARAuAu200Ratios}}} & $ 0 - 10 $ \% & $1.02\pm0.000\pm0.01$ & $0.97\pm0.026\pm0.10$ & $0.77\pm0.022\pm0.08$\\
\end{tabular}
\end{ruledtabular}
\end{table*}

For a more comprehensive comparison, we focus on published results 
of proton antiparticle to particle ratios at $\sqrt{s_{_{NN}}} =$ 
200 GeV from all RHIC experiments.  
This compilation is shown in Figure \ref{Fig:CompareRatios_graphic} 
for the proton ratios, 
with data given in Table \ref{Table:CuCu200}.  
Within the current level of systematics and available data, 
the colliding system size appears to play only a minor 
role in determining the final like-particle ratios of 
low $\langle p_T \rangle$ pions, kaons and protons, and hence 
also in many of the thermal properties of the collision zone.  
 Also,  the effect of any final-state interactions 
on the antiparticle to particle ratios, likely present in central 
Cu+Cu and Au+Au collisions and absent in p+p and d+Au collisions, 
does not play a significant role.  
It is interesting to note, within the context of 
the present results for the intermediate Cu+Cu system, the 
BRAHMS Collaboration result that the strong agreement of 
particle ratios between the (small) p+p and (large) 
central Au+Au colliding systems seen at midrapidity 
holds out to large rapidity ($y \sim 3$), even as the ratios 
themselves are found to decline \cite{BRAHMSPPRatios}. 
Table \ref{Table:CuCu200} also lists pion and kaon ratio results at RHIC,
which are consistent within uncertainities.

In summary, the first results on antiparticle to particle ratios of 
pions, kaons and protons in Cu+Cu collisions at $\sqrt{s_{_{NN}}} =$ 
62.4 and 200 GeV are reported as a function of centrality.  
No strong dependence on centrality is observed for the Cu+Cu data. 
A detailed comparison of the central Cu+Cu results at 200 GeV to 
results from p+p, d+Au, and central Au+Au collisions at RHIC 
indicates that the antiparticle to particle ratios are, for the 
most part, insensitive to the collision species. 
The final values for the antiparticle to particle ratios of 
pions, kaons and protons appear to be primarily driven by 
the collision energy and, within current systematic uncertainties, 
are largely independent of the colliding system.

%
%
%
%
This work was partially supported by U.S. DOE grants 
DE-AC02-98CH10886,
DE-FG02-93ER40802, 
DE-FG02-94ER40818,  
DE-FG02-94ER40865, 
DE-FG02-99ER41099, and
DE-AC02-06CH11357, by U.S. 
NSF grants 9603486, 
0072204,            
and 0245011,        
by Polish KBN grant 1-P03B-062-27(2004-2007),
by NSC of Taiwan Contract NSC 89-2112-M-008-024, and
by Hungarian OTKA grant (F 049823).



\begin{thebibliography}{99}


\bibitem{phobos2} B.\ B.\ Back \textit{et al.},
Nucl. Instr. Meth. A{\bf 499}, 603 (2003).

\bibitem{phobos1} B.\ B.\ Back \textit{et al.}, 
Nucl. Phys. {\bf A661} 690, (1999).
	
\bibitem{whitepaper} B.\ B.\ Back \textit{et al.}, 
Nucl.\ Phys.\ {\bf A757}, 28 (2005).

\bibitem{phobos130} B.\ B.\ Back \textit{et al.}, 
Phys.\ Rev.\ Lett.\ {\bf 87}, 102301 (2001).

\bibitem{PHOBOSPPRatios} B.\ B.\ Back \textit{et al.}, 
Phys.\ Rev.\ C\ {\bf 71}, 021901 (2005).

\bibitem{phobosdAratios} B.\ B.\ Back \textit{et al.}, 
Phys.\ Rev.\ C\ {\bf 70}, 011901(R) (2004).

\bibitem{PHOBOSAuAu200Ratios} B.\ B.\ Back \emph{et al.}, 
Phys.\ Rev.\ C\ {\bf 67}, 021901 (2003).

\bibitem{Garcia:OctVertexPaper} E.\ Garcia \textit{et al.}, 
Nucl. Instr. Meth. A{\bf 570}, 536 (2007).

\bibitem{phobosCuCu_pT} B.\ Alver \textit{et al.}, 
Phys. Rev. Lett. {\bf 96}, 212301 (2006).

\bibitem{phobos:tracklet200_20} B.\ B.\ Back \emph{et al.}, 
Phys.\ Rev.\ C\ {\bf 70}, 021902(R) (2004).

\bibitem{phobos:dAu_dNdEta}B.\ B.\ Back \emph{et al.}, 
Phys.\ Rev.\ C\ {\bf 72}, 031901(R) (2005).

\bibitem{GEANT} Geant Detector Simulation Tool v. 3.21, CERN (default parameters).

\bibitem{HIJING} M.\ Gyulassy \emph{et al.}, 
Phys.\ Rev.\ D \textbf{44}, 3501 (1991).

\bibitem{PHENIXAuAu130lambda} K.\ Adcox \emph{et al.}, 
Phys.\ Rev.\ Lett. {\bf 89}, 092302 (2002).

\bibitem{STARAuAu130pbarp}  C.\ Adler \emph{et al.}, 
Phys.\ Rev.\ Lett.\ {\bf 86}, 4778 (2001).

\bibitem{PHOBOS62AuAuPID} B.\ B.\ Back \emph{et al.}, 
Phys.\ Rev.\ C {\bf 75}, 024910 (2007).

\bibitem{PHOBOSdAuNPart}  B.\ B.\ Back \emph{et al.}, 
Phys.\ Rev.\ C {\bf 72}, 031901 (2005).

\bibitem{BRAHMSPPRatios} I.\ G.\ Bearden \textit{et al.},
Phys. Lett. B {\bf 607}, 42-50 (2005).

\bibitem{STARPPRatios}  B.\ I.\ Abelev \emph{et al.}, 
Phys.\ Rev.\ C {\bf 75}, 064901 (2007), and private communication.

\bibitem{PHENIXAuAu200Ratios} S.\ S.\ Adler \emph{et al.}, 
Phys.\ Rev.\ C\  {\bf 69}, 034909 (2004).

\bibitem{BRAHMSAuAu200Ratios} I.\ G.\ Bearden \textit{et al.},
Phys.\ Rev.\ Lett.\ {\bf 90}, 102301 (2003).

\bibitem{STARAuAu200Ratios} G.\ van Buren \emph{et al.}, 
Nucl.\ Phys.\ {\bf A715}, 129c (2003).

\end{thebibliography}
\end{document}